\documentclass[%
 reprint,
 amsmath,amssymb,
prb,
]{revtex4-1}

\usepackage{graphicx}
\graphicspath{{Pictures/}}
\DeclareGraphicsExtensions{.pdf,.png,.jpg}
\usepackage{dcolumn}
\usepackage{bm}

\begin{document}

\preprint{APS/123-QED}

\title{Size effect on the hysteresis characteristics of a system of interacting core/shell nanoparticles}

\author{Afremov Leonid}

\email{afremov.ll@dvfu.ru}
\affiliation{
	Physics Department, Far Eastern Federal University,Vladivostok, Russia.
}
\author{Anisimov Sergei}
\affiliation{
	Physics Department, Far Eastern Federal University,Vladivostok, Russia.
}
\email{ahriman25@gmail.com}
\author{Iliushin Ilia}
\affiliation{
	Physics Department, Far Eastern Federal University,Vladivostok, Russia.
}
\email{futted@gmail.com}
\author{Qiang You}
\affiliation{
	Please insert your university name
}
\email{and e-mail}

\date{\today}

\begin{abstract}
We have developed a model for the interacting сore/shell nanoparticles, which we used to analyze the dependence of the coercive field $H_c$, the remanent saturation magnetization $M_rs$ and the saturation magnetization $M_s$ on the interfacial exchange interaction between the core and the shell, the size of the nanoparticles and their interaction for Fe/Fe$_3$O$_4$ nanoparticles have been carried out. It has been shown that the hysteresis characteristics increase together with the size of nanoparticles. $H_c$ and $M_rs$ are changing nonmonotonic when the constant interfacial exchange interaction changes from negative to positive values. In the system of core/shell nanoparticles, magnetic interaction results in  $H_c$ and $M_rs$ dropping, which was confirmed by experiments.
\begin{description}

\item[Usage]
Secondary publications and information retrieval purposes.
\item[PACS numbers]
May be entered using the \verb+\pacs{#1}+ command.
\item[Structure]
You may use the \texttt{description} environment to structure your abstract;
use the optional argument of the \verb+\item+ command to give the category of each item. 
\end{description}
\end{abstract}

\pacs{Valid PACS appear here}
\maketitle


\section{\label{sec:level1}Introduction}

Core/shell nanoparticles, where core and shell can represent different combinations of magnetic materials (magnetic / magnetic), occupy a special place amongst nanoparticles with various combinations of core and shell compositions \cite{Ghosh,Walid,Kim,Seyedeh,Lavorato,Yujun,Pana,Tamer,Cho,Soares,Chen,Bahma,Wen,Bigot,Teng,Zeng}. Modern synthesis technologies of core/shell nanoparticles allow one to create nanoparticles such as FePt/Fe$_3$O$_4$ and FePt/CoFe$_2$O$_4$ \cite{Seyedeh}, CoFe$_2$O$_4$/CoFe$_2$\cite{Soares}, Fe/Fe$_3$O$_4$\cite{Kaur}, Fe/ZrO$_2$ \cite{Girija} and Mn/Mn$_3$O$_4$\cite{German}. This list can be supplemented with magnetic nanoparticles coated with a gold shell, such as Fe$_3$O$_4$/Au\cite{Tamer,Lim}, Co/Au\cite{Yujun}, Fe/Au\cite{Cho}, Ni/Au \cite{Dong}, because at sizes of less than 15 nm, the otherwise diamagnetic gold becomes ferromagnetic\cite{Chen,Girija,Skoro,Lopez,Nogues}. The field of applications of these magnetic nanoparticles is quite wide, spanning biomedicine \cite{Cho,Chen,Lim,Dong,Lopez,Ian}, catalysis\cite{Kim,Lim}, electronics\cite{Seyedeh,Kaur,Lopez}, and the creation of nanocomposite materials and films\cite{Teng}.

The magnetic properties of nanoparticles (magnetic/magnetic) significantly depend on the methods of their synthesis\cite{Seyedeh,Pana,Tamer,Cho,Dong,Skoro,Ian,Silva}, on the size of the core, the shell thickness and shape \cite{Lavorato,Pana,Dong,Zeng2,Oscar}. The coercive field $H_c$ \cite{Yujun,Ian,Kaur,Bahma,Wen}, the saturation magnetization $M_s$\cite{Lavorato,Kaur}, the remanent saturation magnetization $M_{rs}$\cite{Lavorato,Kaur} and the blocking temperature $T_b$\cite{Tamer,Bahma,Dong,Chen,Zeng2} of the core/shell particles substantially changed with decreasing size. For example, it has been shown experimentally\cite{Zeng2}, that the coercive field  $H_c$  of the bimetallic FePt/Fe$_3$O$_4$ or FePt/CoFe$_2$O$_4$ nanoparticles decreases with a growing volume fraction of the magnetite or Co-ferrite shell $f_s$. The same dependence of $H_c$ on the thickness of the shell of the core/shell nanoparticles CoFe$_2$O$_4$/CoFe$_2$ was observed by the authors of reference\cite{Soares}. An experimental study on the size effect of the nanoparticle size on $H_c$ and the ratio of $M_{rs}/M_s$ of the nanoparticles CoO/CoFe$_2$O$_4$ has been presented in reference \cite{Lavorato}. It has been shown, that at $T=5$K the size inrease of the nanoparticle leads to a decrease of the coercive field and a growth of the ratio $M_{rs}/M_s$, and the blocking temperature $T_b$. At the same time at the room temperature $T=300$K is increased. A similar increase of the hysteresis characteristics of the Fe/Fe$_3$O$_4$ core/shell nanoparticles that is observed at the room temperature as has been shown in a sufficiently consistent and detailed study\cite{Kaur}.

Please note that decreasing the size of the nanoparticles leads to a decrease in the height \cite{Trohidou}, but the hysteresis characteristics as well ($H_c$, $M_s$, $M_{rs}$)\cite{Nogues,Trohidou}.

We cannot neglect the influence of the magnetic interactions between particles, caused by their mutual arrangement, on their hysteresis characteristics\cite{Zeng2,Kaur,Trohidou}. For example, for a uniform distribution of the $N$ magnetic grains of size $a$ in a nonmagnetic matrix, the field produced by one of the particles on its adjacent ones can be estimated as $H_m=2NM_sa^3/R^3=2cM_s$ where $R$ -- the average distance between the particles, $c=Na^3/R^3$ -- volume concentration of magnetic material in the system. For magnets with a high spontaneous magnetization $M_s\sim500-1500$ G at $c\sim0.1-0.3$ the magnetostatic interaction can have a significant effect on the magnetization of the core/shell nanoparticles.

Theoretical studies of the magnetic properties of core/shell nanoparticles are mainly based on Monte Carlo simulations\cite{Oscar,Vasilakaki,Wu,Kechrakos,Hu}. An exception is a method based on micromagnetic simulation. For example, in \cite{Trohidou}, micromagnetic simulation of magnetization reversals of nanoparticles has been carried out. The flaw of this simulation is that it neglects the thermal fluctuations that limit the application of theoretical results for those particles with volumes lower than the blocking volume. The Monte Carlo method does not have this flaw. A study of the coercive field and the remanent saturation magnetization of core/shell nanoparticles with a ferromagnetic core on the particle size and the thickness of the antiferromagnetic/ferrimagnetic disordered shell has been carried out using this method\cite{Kechrakos}. Investigating the effects of the magnetostatic interaction between the core/shell nanoparticles on their coercivity $H_c$ and blocking temperature $T_b$, conducted using Monte Carlo simulation\cite{Trohidou} showed that a growth of the interparticle interaction lead to the decrease of the $H_c$ and the increase of $T_b$. Similar results are reported in reference\cite{Kechrakos}, where a Monte Carlo simulation was conducted within the Meiklejohn-Bean model. Another way to study the magnetic interactions of nanoparticles is represented in references\cite{Lim,Vasilakaki,Kechrakos,Manakov}. This approach is based on the assumption of a random field interaction of the nanoparticle with all particles of the system. In references\cite{Lim,KechrakosManakov} a method for designing the distribution function of the random fields of the particle interaction that does not limit the type of this interaction has been presented (magnetostatic, direct exchange, RKKY or any other). Moreover, the distribution function essentially depends not only on the location of the nanoparticles (dimension ensemble)\cite{Kechrakos}, but also on their concentration c. Thus, according to references\cite{Vasilakaki,Kechrakos}, at a low volume concentration of dipole-dipole interacting magnetic nanoparticles ($c<0.1$), randomly distributed in a nonmagnetic matrix, the random fields of the magnetostatic interactions h are distributed according to the Cauchy’s law. The distribution of the fields of interactions is normal at $c>0.1$. 

In this work we present a study on the dependence of the coercive field, the saturation magnetization and the remanent saturation magnetization of the core/shell nanoparticles on their size, geometry of core and shell, interaction between the core and the shell, and interparticle magnetic interaction. It has been carried out in the framework of our model of core/shell nanoparticles.

\section{\label{sec:level1}Experiment}

In the current study, five different sizes of iron-iron oxide core-shell NPs were used to compare the theoretical model. These NPs were synthesized using a nanocluster deposition system: a combination of magnetron sputtering and gas aggregation technique. The structural and magnetic properties of core-shell NPs are essentially determined by a ratio of metal core and oxide shell which is highly dependent on cluster size.

The NPs size can be varied by adjusting several parameters such as the ratio of argon (Ar) to helium (He) gas, the power supplied to magnetron sputtering, the aggregation length and the temperature of the aggregation region. The deferent studied NP sizes were achieved by changing the Ar to He gas ratio and keeping the other parameters such as power, aggregation length and temperature constant.  Energetic metal atoms were sputtered from the target cooled by water using the mixture of gases (Ar and He) leading to the nucleation of clusters. The pressure difference allowed the clusters to travel from the aggregation chamber to the deposition chamber. The nucleation and growth of clusters ceases after expansion through a nozzle. A small amount of oxygen (2 sccm) in the deposition chamber reacts with zero-valent crystal iron NPs and forms a protective shell of oxide on NPs before it lands softly onto the surface of silicon Si (100) substrate at room temperature. The detailed deposition method is found in our previous papers\cite{Qiang,Kaur,Kaur2}.

Several measurement techniques have been used to analyze the samples with respect to their corresponding structural, elemental, chemical, and magnetic properties. Rigaku D/MAX RAPID II microdiffractometer (G-XRD) (Cr K$\alpha$, $\lambda=2.2897$\AA) operating at 35 kV and 25 mA was employed to study the crystallographic phase, composition, and average size of crystallite at room temperature. The morphology of individual NC was analyzed using high resolution transmission electron microscopy (HRTEM), a JEOL JEM 2010 microscope with a LaB$_6$ filament operating at 200 kV equipped with a slow-scan CCD camera. The microstructures of different NPs were examined using a helium ion microscope (HIM) from Orion Plus, Carl Zeiss SMT, Peabody, MA at an operating voltage of 30 kV and a probe current of a few pico-amperes, which has a higher surface sensitivity, a better spatial resolution, a larger depth of field, and a higher image contrast compared to scanning electron microscopy. Magnetic properties were studied at room temperature as well as at low temperature for different NCs using a vibrating sample magnetometer (DMS 1660) and a Physical Property Measurement System (PPMS with ACMS option, Quantum Design, San Diego, CA), respectively. The complete characterization and analysis is reported in our previous work\cite{Qiang,Kaur,Kaur2}.

\section{\label{sec:level1}Model of the core/shell nanoparticles}

\begin{figure}
	\center{\includegraphics[scale=0.4]{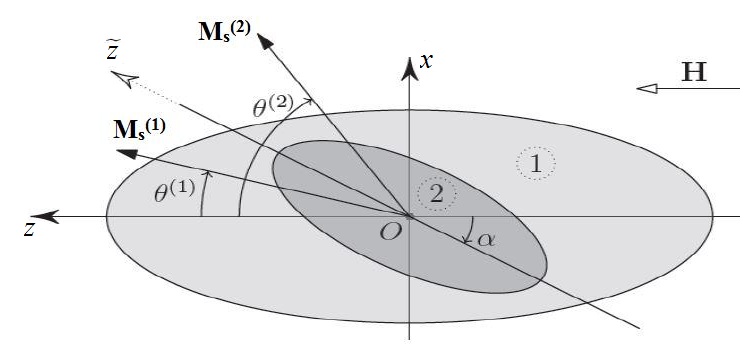}}
	\caption{Illustration for the model of the core/shell nanoparticle}
	\label{fig:fig1}
\end{figure}

\begin{enumerate}
	\item We assume a uniformly magnetized ellipsoidal magnetite nanoparticle (phase (1)) of volume $V$ with an elongation $Q$ containing a uniformly magnetized ellipsoidal iron core (phase (2)) of volume $v = \varepsilon V$ and elongation q. The long axis of the second phase makes an angle $\alpha$ with the long axis of the nanoparticles oriented along the axis $Oz$ (see. Fig.\ref{fig:fig1}).
	\item It is believed that the anisotropy axis is parallel to the long axis of the ferromagnetic nanoparticle and the core.
	\item The vectors of the spontaneous magnetization of both phases $M_s^{(1)}$ and $M_s^{(2)}$ are located in the plane $xOz$ containing the long axis of the magnetic phases and make the angles $\vartheta^{(1)}$ and $\vartheta^{(2)}$ with the axis $Oz$, respectively.
	\item An external magnetic field $H$ is applied along the axis $Oz$. 
\end{enumerate}

\subsection{\label{sec:level2}Magnetic states of the core/shell nanoparticles}

It is known (see, e.g., reference\cite{Neel}), that the magnetic state of a nanoparticles is significantly depends on its size and temperature. Moreover, for a given volume of nanoparticles one can always indicate a blocking temperature $T_b$ below which the thermal fluctuations have almost no influence on the magnetic moment $m$. For  $T>T_b$, the nanoparticles move to a superparamagnetic state with an average magnetic moment $\langle m \rangle=0$. 
Let us start by studying the magnetic state of nanoparticles with a blocked magnetic moment.

\subsubsection{Magnetic state of the uniaxial core/shell nanoparticles}

The energy $E$ of a nanoparticle in an external field $H$ can be represented as the sum of:

\begin{itemize}
	\item the crystallographic anisotropy constant: 

\begin{equation}\label{eq:eq1}
	\begin{matrix}
		E_A=-\frac{1}{4}\{(M_s^{(1)})^2k_A^{(1)}(1-\varepsilon)\cos{2\vartheta^{(1)}}+\\(M_s^{(2)})^2k_A^{(2)}\varepsilon\cos{2(\vartheta^{(2)}-a)}\}V
	\end{matrix}
\end{equation}

	\item the interaction energy of the magnetic moment with its own magnetic field, which, in accordance to Appendix I can be represented as: 

\begin{equation}\label{eq:eq2}
	\begin{matrix}
		E_m= \{-\frac{(M_s^{(1)})^2}{4}[( (1-2\varepsilon)k_N^{(1)}+\varepsilon k_N^{(2)}\cos{2\alpha})\\\cos{2\vartheta^{(1)}}-\varepsilon k_N^{(2)}\sin{2\alpha}\sin{2\vartheta^{(1)}}]-\frac{(M_s^{(2)})^2}{4}\varepsilon k_N^{(2)}\\\cos2(\vartheta^(2)+\alpha)+\frac{\varepsilon M_s^{(1)}M_s^{(2)}}{3}[\frac{3}{2}k_N^{(2)}\sin{2\alpha}\sin(\vartheta^{(1)}+\\\vartheta^{(2)})+(k_N^{(2)}-k_N^{(1)})(\sin\vartheta^{(1)}\sin\vartheta^{(2)}-\\2\cos\vartheta^{(1)}\cos\vartheta^{(2)})]\}V
	\end{matrix}
\end{equation}
	
	\item the exchange interaction energy across the border: 

\begin{equation}\label{eq:eq3}
	E_{ex}=-\frac{2A_{in}}{\delta}\cos(\vartheta^{(1)}-\vartheta^{(2)})s
\end{equation}

	\item Zeeman's energy: 
	
\begin{equation}\label{eq:eq4}
E_H=-H[(1-\varepsilon)M_s^{(1)}\cos \vartheta^{(1)}+\varepsilon M_s^{(2)}\cos\vartheta^{(2)}]
\end{equation}
\end{itemize}

In the formulae (\ref{eq:eq1})-(\ref{eq:eq5}) $k_A^{(1,2)}=K_1/(I_s^{(1,2)})^2, k_N^{(1,2)}$ -- dimensionless anisotropy constants and  shape anisotropy of the phases, respectively, $K_1^{(1,2)}$ -- the first anisotropy constants of the phases, $V$ -- the volume of a nanoparticle, $s$ -- the surface area between the phases, $\varepsilon$ --the ratio of the inclusion's volume to the total volume of the particle, $A_{in}$ -- the interfacial exchange interaction constant, $\delta$ -- the width of the transition region with the order of a lattice constant. Please note that the shape anisotropy constant $k_N=2\pi (1-3N_z)$ is expressed through the demagnetizing factor along the long axis $N_z$, and depends only on the elongation of the ellipsoid $q$: $N_z=[q\ln(q+\sqrt{q^2-1})]-\sqrt{q^2-1}]/(q^2-1)^{3/2}$.

An analysis of the magnetic states is performed by minimizing the total free energy $E=E_A+E_m+E_{ex}+E_H$ of a nanoparticle:

\begin{equation}\label{eq:eq5}
	\begin{matrix}
		E=\{ -\frac{(M_s^{(1)})^2}{4}\mathcal{K}^{(1)}\cos2(\vartheta^{(1)}-\delta^{(1)}) -\frac{(M_s^{(2)})^2}{4}\mathcal{K}^{(2)}\\\cos2(\vartheta^{(2)}-\delta^{(2)})+M_s^{(1)}M_s^{(2)}[-\mathcal{U}_1\sin\vartheta^{(1)}\sin\vartheta^{(2)}+\\\mathcal{U}_2\cos\vartheta^{(1)}\cos\vartheta^{(2)}+\frac{1}{2}\varepsilon k_N^{(2)}\sin2\alpha\sin(\vartheta^{(1)}+\vartheta^{(2)}) ]\\-H[(1-\varepsilon)M_s^{(1)}\cos\vartheta^{(1)}+\varepsilon M_s^{(2)}\cos\vartheta^{(2)}]\}V
	\end{matrix}
\end{equation}

where effective anisotropy constant $\mathcal{K}^{(1,2)}$ and  the position of effective axes phase $\delta^{(1,2)}$ were calculated:

\begin{equation}\label{eq:eq6}
	\begin{matrix}
		\mathcal{K}^{(1)}=[\left((1-\varepsilon)k_A^{(1)}+(1-2\varepsilon)k_N^{(1)}+\varepsilon k_N^{(2)}\cos2\alpha\right)^2\displaybreak[4]\\+(\varepsilon k_N^{(2)}\sin2\alpha)^2]^{\frac{1}{2}}
	\end{matrix}
\end{equation}

\begin{equation}\label{eq:eq7}
	\tan(2\delta^{(1)})=-\frac{\varepsilon k_N^{(2)}\sin2\alpha}{(1-\varepsilon)k_A^{(1)}+(1-2\varepsilon)k_N^{(1)}+\varepsilon k_N^{(2)}\cos2\alpha}
\end{equation}

\begin{equation}\label{eq:eq8}
	\mathcal{K}^{(2)}=\varepsilon\sqrt{(k_A^{(2)})^2+(k_N^{(2)})^2+2k_A^{(2)}k_N^{(2)}\cos4\alpha}
\end{equation}

\begin{equation}\label{eq:eq9}
	\tan(2\delta^{(2)})=-\frac{k_N^{(2)}-k_A^{(2)}}{k_N^{(2)}+k_A^{(2)}}\tan2\alpha
\end{equation}

The constants of interfacial interaction $\mathcal{U}_1$ and $\mathcal{U}_2$ expressed in terms of constant exchange and magnetostatic interaction phases:

\begin{equation}\label{eq:eq10-1}
	\begin{matrix}
		\mathcal{U}_1=\varepsilon\left(\frac{(k_N^{(1)}-k_N^{(2)})}{3}+\frac{2sA_{in}}{\nu\delta M_s^{(1)}M_s^{(2)}}\right),
			\end{matrix}
			\end{equation}
			
\begin{equation}\label{eq:eq10-2}
	\begin{matrix}
		\mathcal{U}_2=\varepsilon\left(\frac{2(k_N^{(1)}-k_N^{(2)})}{3}-\frac{2sA_{in}}{\nu\delta M_s^{(1)}M_s^{(2)}}\right),
	\end{matrix}
\end{equation}

\begin{equation}\label{eq:eq11-1}
	\begin{matrix}
		\frac{\mathcal{K}^{(1)}}{2}\sin2(\vartheta^{(1)}-\delta^{(1)})
		-j[\mathcal{U}_1\cos\vartheta^{(1)}\sin\vartheta^{(2)}+\\\mathcal{U}_2\sin\vartheta^{(1)}\cos\vartheta^{(2)}
		+\frac{3}{2}\varepsilon k_N^{(2)}\sin2\alpha\cos(\vartheta^{(1)}+\vartheta^{(2)})]
		\\+h(1-\varepsilon)\sin\vartheta^{(1)}=0,
	\end{matrix}
\end{equation}

\begin{equation}\label{eq:eq11-2}
	\begin{matrix}		
		\frac{j\mathcal{K}^{(2)}}{2}\sin2(\vartheta^{(2)}-\delta^{(2)})
		-[\mathcal{U}_1\sin\vartheta^{(1)}\cos\vartheta^{(2)}+\\\mathcal{U}_2\cos\vartheta^{(1)}\sin\vartheta^{(2)}
		+\frac{3}{2}\varepsilon k_N^{(2)}\sin2\alpha\cos(\vartheta^{(1)}+\\\vartheta^{(2)})]+h\varepsilon\sin\vartheta^{(2)}=0,
	\end{matrix}
\end{equation}

where $j=\frac{M_s^{(1)}}{M_s^{(2)}}, h=\frac{H}{M_s^{(1)}}$.

The system of equations (\ref{eq:eq11-1})-(\ref{eq:eq11-2}) together with the terms of the minimum energy (\ref{eq:eq5})

\begin{equation}\label{eq:eq12}
	\frac{\partial^2E}{\partial\vartheta^{(1)2}}>0,\quad\frac{\partial^2E}{\partial\vartheta^{(1)2}}\frac{\partial^2E}{\partial\vartheta^{(2)2}}-\left(\frac{\partial^2E}{\partial\vartheta^{(1)}\partial\vartheta^{(2)}}\right)^2>0
\end{equation}

It defines the basic and metastable states of the magnetic moments of nanoparticles phases.

\subsubsection{Magnetic state of multi-core/shell nanoparticles}

We assume that the magnetic phases of the nanoparticles are represented by crystals of cubic symmetry. To satisfy the conditions described in paragraph 2 of the model of core/shell nanoparticles, we align one of their crystallographic directions [100], [010] and [001] with the long axis of the phase, if the anisotropy constant of the first phase is $\tilde{k}_{A1}^{(1,2)}>0$. In case of $\tilde{k}_{A1}^{(1,2)}<0$, we align it with the direction [111].

We use the condition of magnetic uniaxiality of the multiaxial crystal\cite{Afremov1,Afremov2}, the essence of which being that at a certain elongation, its shape anisotropy prevails over the crystalline anisotropy. The process of magnetization for the particles is similar to the magnetization of uniaxial particles with constant anisotropy:

\begin{equation}\label{eq:eq13}
k_A^{(1,2)}=\begin{cases}\tilde{k}_{A1}^{(1,2)}, & \tilde{k}_{A1}^{(1,2)}>0,\\
\frac{1}{3}\tilde{k}_{A2}^{(1,2)}, & \tilde{k}_{A1}^{(1,2)}<0.
\end{cases}
\end{equation}

where $\tilde{k}_{A1}=K_1/(M_s)^2, \tilde{k}_{A2}=K_2/(M_s)^2$ -- dimensionless anisotropy constants, $K_1, K_2$ -- first and second magnetic anisotropy constant of a cubic crystal, respectively.

Please note that for many materials, the condition of magnetic uniaxiality\cite{Afremov1,Afremov2} [39,40] is fulfilled for nanoparticles with low elongation. Examples include iron $k_N=\tilde{k}_{A1}$ at $q\approx1.03$ and magnetite at $q\approx1.09$.

Hence, to determine the magnetic states of multiaxial particles with $k_N>\tilde{k}_{A1}$, the formulae (\ref{eq:eq5})--(\ref{eq:eq11-2}) can be used after replacing the anisotropy constant $k_A^{(1,2)}$ using equation (\ref{eq:eq13}).

\subsubsection{Ground and metastable magnetic states of core/shell nanoparticles}

Equations (\ref{eq:eq6}) - (\ref{eq:eq12}) show that the states of core/shell nanoparticles with the desired magnetic core and shell materials, e.g. Fe/Fe$_3$O$_4$, are determined by the size and shape of the nanoparticle and its core. The calculations carried out using equations (\ref{eq:eq11-1})-(\ref{eq:eq12}) showed that at $Q\ge1$ and $q\ge1$, for all values of angle $\alpha$ between the long axes of the nanoparticles and the core, there are four or two magnetic states available for the nanoparticles. The states differ in the relative orientation of the magnetic moments of the phases. For example, at $\alpha=0$, the magnetic moments of the core and shell are oriented parallel ($\uparrow\uparrow,\downarrow\downarrow$) or antiparallel ($\uparrow\downarrow,\downarrow\uparrow$) to each other \cite{Afremov3,Afremov4}. the increase of $\alpha$ leads to a growth of the deviation of the magnetic moments from the axis $Oz$ (see Fig. \ref{fig:fig1}) in each of the four states:

\begin{itemize}
	\item \textit{in the first} ''($\nwarrow\nearrow$)-state'', the magnetic moments of both phases make sharp angles ($-\pi/2<\vartheta^{(1)}<\pi/2, -\pi/2<\vartheta^{(2)}<\pi/2$) with the axis $Oz$;
	\item \textit{in the second} ''($\nearrow\searrow$)-state'', the magnetic moments of both phases make angles $-\pi/2<\vartheta^{(1)}<\pi/2, -3\pi/2<\vartheta^{(2)}<-\pi/2$ with the axis $Oz$;
	\item \textit{the third} ''($\swarrow\searrow$)'' and \textit{the fourth} ''($\swarrow\nwarrow$)'' states are the inversed first and second states, respetively.
\end{itemize}

Moreover, if the magnetostatic interaction between the phases dominates over the exchange interaction $\mathcal{U}_2>0$, the second and fourth states are stable, whilst the first and third states are metastable, since the free energy of a particle in these states is less than in the first and third. Otherwise ($\mathcal{U}_2<0$), the first and third states are stable.

\begin{figure}
	\center{\includegraphics[scale=0.9]{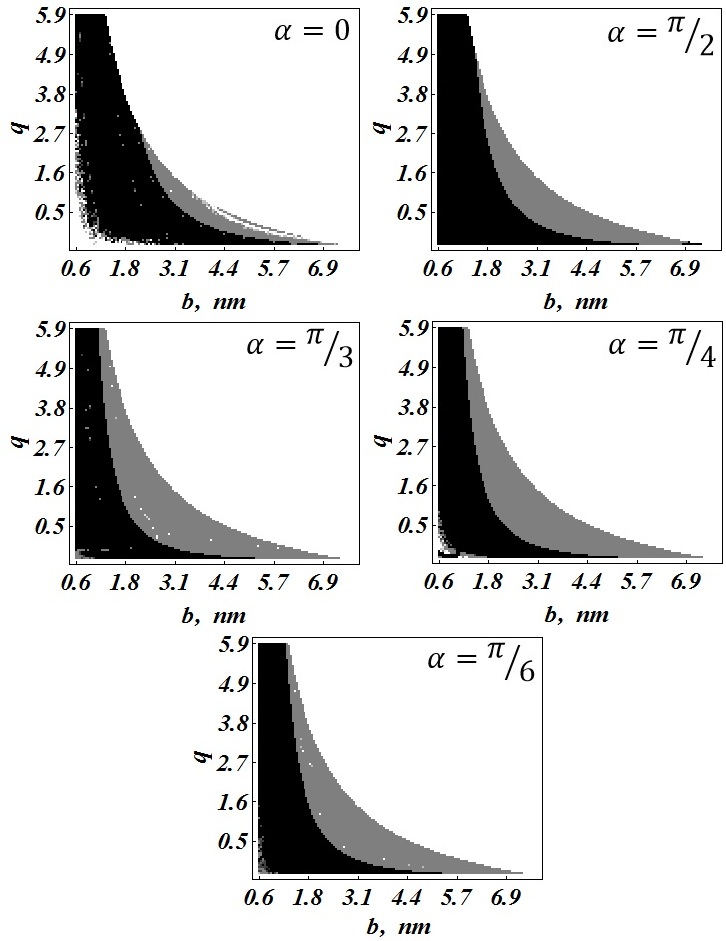}}
	\caption{A diagram$\{b,q\}$ for different $\alpha=0$, $\pi/6$, $\pi/4$, $\pi/3$, $\pi/2$.}
	\label{fig:fig2}
\end{figure}

The dependence of the magnetic states of the nanoparticles, with given sizes of the small semiaxis $B$ and the elongation $Q$, on the geometric characteristics of the core (the size of a small semiaxis $b$ and elongation $q$) is conveniently represented by the diagram \{$b,q$\} shown in fig. 2.

Each point of the diagrams \{$b,q$\} is a nanoparticle, the core of which has a size $b$ and elongation $q$. Black points represent those nanoparticles that are in one of the four ground or metastable states described above. Grey points represent nanoparticles in one of the two equilibrium ground states. 

To plot diagrams\{$b,q$\}, we used a ratio between the short semiaxes of the core ellipsoid $b$ and the nanoparticle $B$ of ($b\le B$), and the restriction $qb\le R(\alpha)$ for the long axis of the core due to the size of the nanoparticles $R(\alpha)=QB/\sqrt{\cos^2\alpha+Q^2\sin^2\alpha}$ along a line that coincides with the long axis of the core. The abovementioned equations set the limitations on the choice of $q$ and $b$:

\begin{equation}\label{eq:eq14}
\frac{b}{B}\leqslant 1, q\frac{b}{B}\leqslant \frac{Q}{\sqrt{\cos^2\alpha+Q^2\sin^2\alpha}}
\end{equation}

Please note that the maximum number of the equilibrium states of a two-phase core/shell nanoparticles is twice the number of "easy axes". Therefore, a spherical core/shell nanoparticle with a spherical core, the magnetic phases of which are represented by a material with cubic symmetry and which does not satisfy the uniaxial magnetic condition\cite{Afremov1,Afremov2}, may be in one of six or eight states.

\subsubsection{Thermal fluctuation effects on the magnetic states of core/shell nanoparticles}

An increase of the temperature $T$ or a decrease of the grain volume $V$ leads to an increase of the thermal fluctuations of the magnetic moments of the phases of the nanoparticles and, therefore, to significant changes of the magnetic properties of the system. This is due to changes in the ratio of the thermal energy $kT$ to the potential barrier $E_{ik}=E_{ik max}-E_{i min}$ separating $i$-th and $k$-th states ($E_{ik max}$ -- the smallest of the maximal values of the energy separating $i$-th and $k$-th states, $E_{i min}$ -- energy of the $i$-th equillibrium state). Since the transition probability from state $i$ to $k$ is $P_{ik}\sim \exp(-E_{ik}/k_BT)$, we consider the transition frequency $W_{ik}=f_0\exp(-E_{ik}/k_BT)$, where $f_0=10^9\div10^{10}s^{-1}$ \cite{Neel}. Detailed barrier $E_{ik}$ calculations can be found in reference\cite{Afremov2} and the results of this research can be found in Appendix II. 

We define the equilibrium state population $l$ of the equilibrium states of the system of core/shell nanoparticles via a normalized population vector $N(t)=\{n_1(t),n_2(t),\cdots,n_l(t)\}$. The components of the vector $N(t)$ can be defined as the probabilities of finding a nanoparticle in one of the equilibrium states. If the system of core/shell nanoparticles is in a non-equilibrium state with $N(0)=\{n_1(0)=n_{01},n_2(0)=n_{02},\cdots,n_l(0)=n_{0l}\}$, then according to reference \cite{Afremov2}, the transition back to the equilibrium state can be described by the following equations: 

\begin{equation}\label{eq:eq15}
\frac{dn_i(t)}{dt}=\sum_{k\neq i}^{l}\left(-W_{ik}n_i(t)+W_{ki}n_k(t)\right)
\end{equation}

Expressing $\Sigma_{i=1}^l n_i(t)=1$ from the normalization condition, the population of the $l$-th state is

\begin{equation}\label{eq:eq16}
n_l(t)=1-\sum_{i=1}^{l-1}n_i(t)
\end{equation}

Excluding $n_l(t)$ out of (\ref{eq:eq15}), equation can be written as followed:

\begin{equation}\label{eq:eq17}
\frac{dn(t)}{dt}=\mathcal{W}n(t)+\mathcal{V},
\end{equation}

Here, the matrix elements of the matrixes $\mathcal{W}, \mathcal{V}$, and the vectors $n(t)$ and $n(0)$ can be expressed in terms of $W_{ik}$ and $N(t)$, respectively:

\begin{equation}\label{eq:eq18}
\begin{matrix}
\mathcal{W}_{ik}=\begin{cases}-\sum_{j\neq i,j=1}^{l}W_{ij}-W_{li}, & i=k \\
W_{ki}-W_{li}, & i\neq k, \end{cases}\\
\mathcal{V}=\begin{pmatrix} W_{l1}\\W_{l2}\\...\\W_{l-1}\end{pmatrix},
n(t)=\begin{pmatrix} n_1(t)\\n_2(t)\\...\\n_{l-1}(t)\end{pmatrix},
n(0)=\begin{pmatrix} n_{01}\\n_{02}\\...\\n_{0l-1}\end{pmatrix}
\end{matrix}
\end{equation}

A solution of equation (\ref{eq:eq17}) can be represented in vector form via matrix exponential:

\begin{equation}\label{eq:eq19}
n(t)=exp(\mathcal{W}(t))n(0)+\int_{0}^{t}exp(\mathcal{W}(t-\tau))d\tau\mathcal{V}.
\end{equation}

The equations for the energy (\ref{eq:eq5})--(\ref{eq:eq10-2}) and formulae (\ref{eq:eq18}), (\ref{eq:eq19}) allow us to investigate the dependence of the magnetic states of core/shell nanoparticles on their geometric and magnetic characteristics, and such “external” factors as temperature, time and external field.

\subsubsection{Magnetization of a system of non-interacting core/shell nanoparticles}

Let us consider a system of N core/shell nanoparticles uniformly distributed in a volume $V_0$. We assume that the nanoparticles of size $B$ are distributed according to the probability $F(B)dB$. According to the model we presented above, the system's magnetization is:

\begin{equation}\label{eq:eq20}
M(t)=\frac{1}{V_0}\int\sum_{i=1}^{l}m_in_i(t,B)F(B)dB,
\end{equation}

where $m_i$ -- magnetic moment projection of the nanoparticle situated in $i$-th state on the external field direction $H$, $n_i(t,B)$ -- population vector components, defined by (\ref{eq:eq19}), $F(B)$ -- size distribution function of the nanoparticles. 

Furthermore, we assume that the long axes of the core and shell of the nanoparticle coincide ($\alpha=0$) and meet the uniaxial magnetic condition \cite{Afremov1,Afremov2}. In this case, (\ref{eq:eq20}) can be written as:

\begin{equation}\label{eq:eq21}
\begin{matrix}
M(t)=c\int[((1-\varepsilon)M_s^{(1)}+\varepsilon M_s^{(2)})(n_1(t,B)-n_3(t,B))+\\((1-\varepsilon)M_s^{(1)}-\varepsilon M_s^{(2)})(n_2(t,B)-n_4(t,B))]F(B)dB.
\end{matrix}
\end{equation}

where $c=NV/V_0$ -- volume concentration of core/shell nanoparticles, $n_4(t,B)=1-n_1(t,B)-n_2(t,B)-n_3(t,B)$.

\subsubsection{Hysteresis characteristics of Fe/Fe$_3$O$_4$ nanoparticles}

The coercive field $H_c$, saturation magnetization $M_s$ and remanent magnetization $M_{rs}$ are determined from the hysteresis loop. Since the system contains nanoparticles with magnetic moments susceptible to thermal fluctuations, nanoparticles with a relaxation time $\tau$ more than the calculation time $t (\tau\ge t)$ contribute to the magnetization. We assume that $t=1$s. Moreover, we take into account the dependence of the magnetization and crystallographic anisotropy of the iron nanoparticles on their size.

Experimental values for the  iron \cite{Matsuura} and magnetite \cite{Caruntu} magnetization have been approximated using the following equations: $M_s^{(Fe)}(B)=-934.28+135.1B-3.14B^2$, $M_s^{(Fe_3O_4)}(B)=60.54+3.86B-0.12B^2)$. The dependence of the crystallographic anisotropy on the size has been determined in a known manner\cite{Gubin}: $K_A=K_V+K_S/B$, where $K_V$ and $K_S$ -- constants of volume and surface anisotropies, respectively, which are $K_V=4.8\cdot10^5$erg/cm$^3$, $K_S=0.04$erg/cm$^2$ for iron, according to references\cite{Graham,Gradmann}, and $K_V=-1.06\cdot10^5$erg/cm$^3$, $K_S=0.029$erg/cm$^2$ for magnetite, according to references \cite{Krupichka,Perez}.

To integrate (\ref{eq:eq21}), the law of lognormal distribution has been used:

\begin{equation}\label{eq:eq22}
F(B)=\frac{1}{B\sqrt{2\pi\sigma^2}}exp\left(-\frac{(\log[B/\langle B\rangle])^2}{2\sigma^2}\right),
\end{equation}

using the  mean size $\langle B\rangle$ and dispersion σ equal to the experimental data shown in work \cite{Kaur}. 

\begin{figure}
	\center{\includegraphics[scale=0.7]{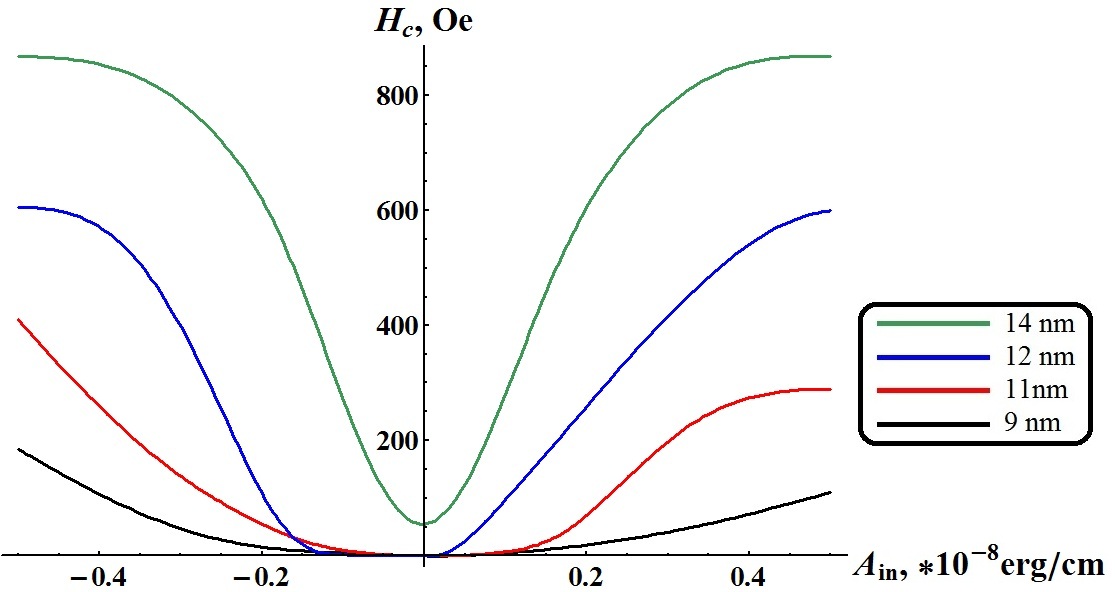}}
	\caption{Dependence of coercive field $H_c$ of the Fe/Fe$_3$O$_4$ nanoparticles of different sizes on the interaction constant $A_{in}$. The shell thickness is $d=2$ nm.}
	\label{fig:fig3}
\end{figure}

\begin{figure}
	\center{\includegraphics[scale=0.7]{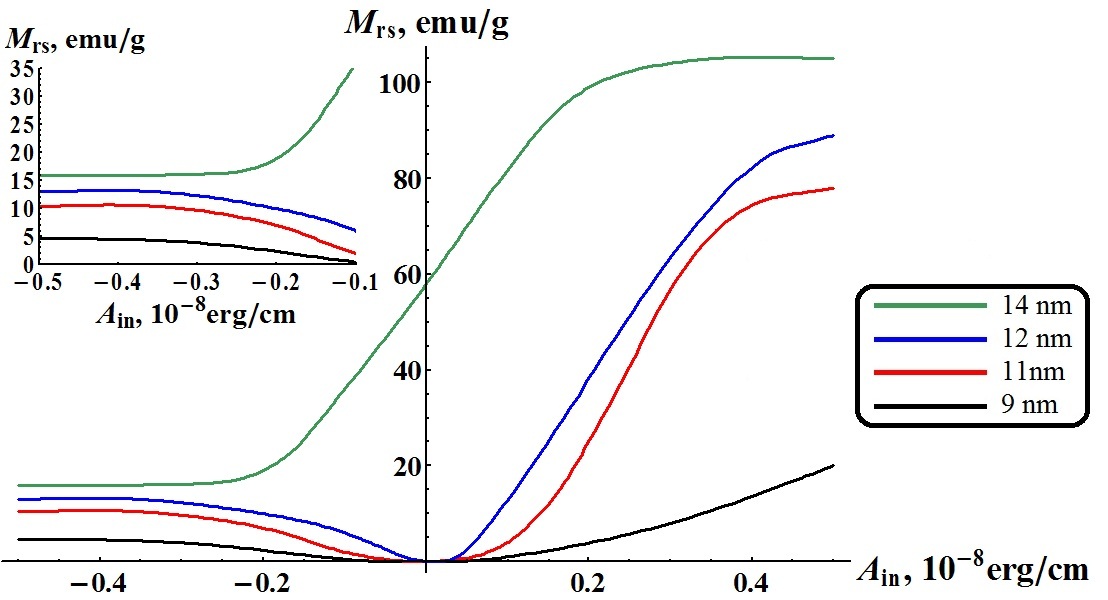}}
	\caption{Dependence of the remanent saturation magnetization $M_{rs}$ of Fe/Fe$_3$O$_4$ nanoparticles of different sizes with a constant shell thickness of $d=2$ nm on the interfacial exchange interaction constant $A_{in}$. The inset shows the dependence of $M_{rs}$ in the area of negative values of $A_{in}$.}
	\label{fig:fig4}
\end{figure}

The results of modeling the hysteresis characteristics of Fe/Fe$_3$O$_4$ nanoparticles of different sizes to the interfacial exchange interaction constant $A_{in}$ are shown on fig. \ref{fig:fig3} and \ref{fig:fig4}. The shell thickness is assumed to be constant at $d=2$ nm and the elongation of the iron core is $q=1.1$.

An increase of the exchange interaction leads to non-monotonic behavior of coercive field $H_c$. This is due to special aspects of the remagnetization of the core/shell nanoparticles. As mentioned in Appendix II, the critical fields for the remagnetization of nanoparticles (equations (11AII) – (16AII)), except for $H_c^{(1\to3)}=H_c^{(3\to1)}$ and $H_c^{(2\to4)}=H_c^{(4\to2)}$ are linearly dependent on the interfacial exchange interaction constant $A_{in}$. This leads to a quadratic dependence of the potential barriers $E_{ik}$ on $A_{in}$ ((1AII)-(10AII)).

The significant exceedance of the remanent saturation magnetization $M_{rs}$ of Fe/Fe$_3$O$_4$ nanoparticles at $A_{in}>0$ over values at $A_{in}<0$ is due to special aspects of the equilibrium states of the core/shell nanoparticles described in 1.3. According to (10), in the area of negative values of $A_{in}$, the second interfacial exchange interaction constant is $\mathcal{U}_2>0$. In this situation, the stable equilibrium state is the second ($\uparrow\downarrow$) or the fourth ($\downarrow\uparrow$) state, where the remanent saturation magnetization is $$M_{rs}(A_{in}<0)\sim M_s^{(\uparrow\downarrow)}=\mid(1-\varepsilon)M_s^{(Fe)}-\varepsilon M_s^{(Fe_3O_4)}\mid.$$
At $A_{in}>0$, the nanoparticle can be in the first ($\uparrow\uparrow$) or the third ($\downarrow\downarrow$) state: $$M_{rs}(A_{in}>0)\sim M_s^{(\uparrow\uparrow)}=(1-\varepsilon)M_s^{(Fe)}-\varepsilon M_s^{(Fe_3O_4)},$$ and this value is higher than $M_{rs} (A_{in}<0)$. Moreover, thermal fluctuations lead to a significantly faster decrease of $M_{rs} (A_{in}<0)$ than $M_{rs} (A_{in}>0)$, since the potential barriers between the third and the first states are higher than those between the second and the fourth states ((3AII), (4AII), (9AII), (10AII)). 

Please note that the vanishing of the coercive field $H_c$ and the remanent magnetization $M_{rs}$ of Fe/Fe$_3$O$_4$ nanoparticles of size $B<12$ nm at low values of the interfacial exchange interaction is due to the transition of the nanoparticles to the superparamagnetic state (fig. \ref{fig:fig3}, \ref{fig:fig4}). Specifically, the decrease of $M_{rs}$ with increasing the interfacial exchange interaction constant at $A_{in}<0$ is caused by a superparamagnetic transition.

\subsubsection{Magnetic interaction in a system of core/shell nanoparticles}

As has been noted in the introduction, for a low volume concentration of the magnetic nanoparticles $c (c<0.1)$, the random fields of the magnetostatic interaction $h$ are distributed according to Cauchy\textsc{\char13}s law\cite{Lim,Nogues}:

\begin{equation}\label{eq:eq23}
W_1(h,M,F)=\frac{1}{\pi}\frac{F}{F^2+[h-\hat{h}(M)]^2}
\end{equation}

where $\hat{h}(M)=(N-8\pi/5)M(H)$ -- mean (the most probable) interaction field. The distribution parameter $B$ (intristic interaction field) and the magnetization of the system of nanoparticles $M(H)$ are defined by the equations:

\begin{equation}\label{eq:eq24}
\begin{matrix}
F=5c\int[((1-\varepsilon)I_s^{(1)}+\varepsilon I_s^{(2)})(n_1(t,h)+n_3(t,h))\\+((1-\varepsilon)I_s^{(1)}-\varepsilon I_s^{(2)})(n_2(t,h)\\+n_4(t,h)]W_1(h,M,F)dh,
\end{matrix}
\end{equation}

\begin{equation}\label{eq:eq25}
\begin{matrix}
M(H)=c\int[((1-\varepsilon)I_s^{(1)}+\varepsilon I_s^{(2)})(n_1(t,H+h)\\-n_3(t,H+h))+((1-\varepsilon)I_s^{(1)}-\varepsilon I_s^{(2)})\\(n_2(t,H+h)-n_4(t,H+h)]W_1(h,M,F)dh,
\end{matrix}
\end{equation}

where $N$ -- demagnetization factor of the system along the external field $H$.

If $c>0.1$, then the distribution of the interaction fields is normal:

\begin{equation}\label{eq:eq26}
W_2(h,M,F)=\frac{1}{\sqrt{\pi F^2}}exp\left(-\frac{(h-\hat{h}(M))^2}{F^2}\right).
\end{equation}

Here $\hat{h}M(H)=NM(H)$, and the magnetization $M(H)$ is defined by (\ref{eq:eq3}) (where necessary to change $W_1(h,M,B)$ to $W_2(h,M,B)$), and the intristic interaction field $B$ is a solution of the following equation \cite{Seyedeh}:

\begin{equation}\label{eq:eq27}
\begin{matrix}
F^2=c\int[((1-\varepsilon)I_s^{(1)}+\varepsilon I_s^{(2)})^2(n_1(t,h)+n_3(t,h))\\+((1-\varepsilon)I_s^{(1)}-\varepsilon I_s^{(2)})^2(n_2(t,h)\\+n_4(t,h))]W_2(h,M,F)dh.
\end{matrix}
\end{equation}

The system of self-consistent equations (\ref{eq:eq24}), (\ref{eq:eq25}), (\ref{eq:eq27}) along with formulae (\ref{eq:eq18}), (\ref{eq:eq19}), which estimate the population vector $n(t,H)$, allows us to calculate the magnetization $M(H)$ of a system of interacting core/shell nanoparticles.

\subsubsection{Magnetic interaction effects on the hysteresis characteristics}

\begin{figure}
	\center{\includegraphics[scale=1]{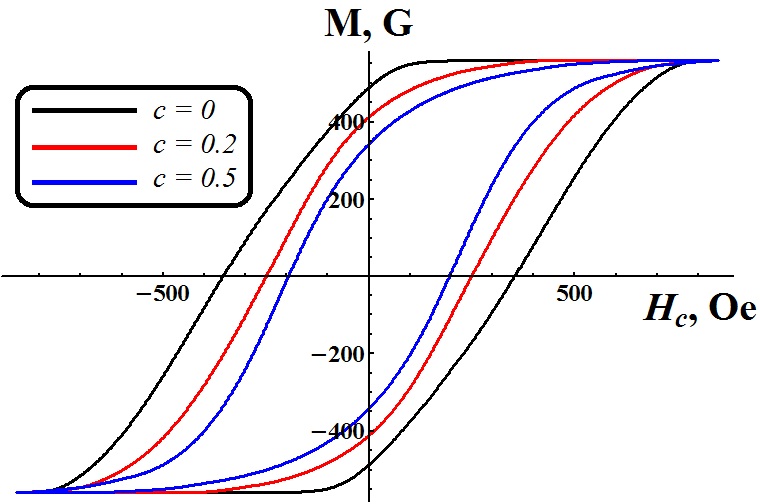}}
	\caption{Dependence of the hysteresis loop of a system of $Fe/Fe_3O_4$ core/shell nanoparticles on the volume concentration $c$. The size of the nanoparticles is 14 nm.}
	\label{fig:fig5}
\end{figure}

\begin{figure}
	\center{\includegraphics[scale=0.45]{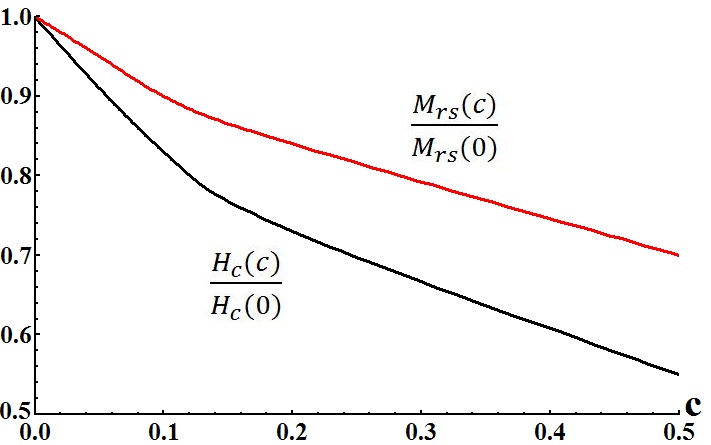}}
	\caption{Dependence of the relative values of the coercive field $H_c(c)/H_c(0)$ and the remanent saturation magnetization $M_{rs}(c)/M_{rs}(0)$ of a system of $Fe/Fe_3O_4$ core/shell nanoparticles with size 14 nm on the volume concentration $c$.}
	\label{fig:fig6}
\end{figure}

The dependence of the hysteresis loop on the intensity of the magnetic interaction in a system of Fe/Fe$_3$O$_4$ nanoparticles is shown in fig. \ref{fig:fig5}. The increase of the volume concentration $c$ of magnetic particles and, respectively, the magnetostatic interaction leads to a “smoothing” of the hysteresis loop, and results in a decrease of the coercive field $H_c$ and the remanent saturation magnetization $M_{rs}$. Fig. \ref{fig:fig6} show the dependences of the relative values of the coercive field $H_c(c)/H_c(0)$ and the remanent saturation magnetization $M_{rs}(c)/M_{rs}(0)$ of a system of Fe/Fe$_3$O$_4$ core/shell nanoparticles with mean size 14 nm on the volume concentration $c$. The decrease of the hysteresis characteristics with the growth of the concentration c of magnetic particles in the sample is due to the chaotization effect of the random interaction field on the distribution of the magnetic moments of the sample.

Please note that the magnetic interaction has a more pronounced effect on the coercive field than on the remanent saturation magnetization, whereas the weak interaction ($c<0.1$) results in a sharper decrease of $H_c$ and $M_{rs}$ than the strong interaction ($c>0.1$).

\begin{figure}
	\center{\includegraphics[scale=0.38]{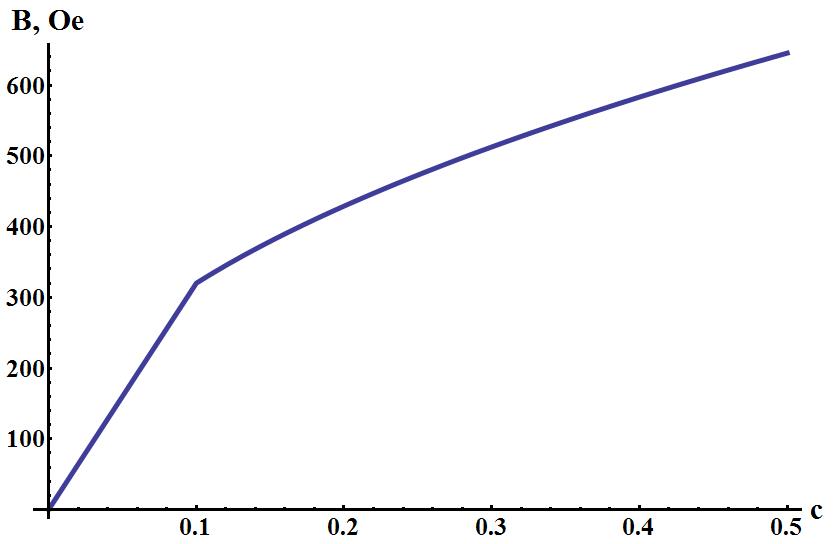}}
	\caption{Dependence of the effective interaction field on the volume concentration $c$ of Fe/Fe$_3$O$_4$ core/shell nanoparticles with size 14 nm.}
	\label{fig:fig7}
\end{figure}

This is due to a faster increase of the effective interaction field $B$ with the growth of the volume concentration of the magnetic nanoparticles at $c<0.1$, than at $c>0.1$ (fig. \ref{fig:fig7}).

\begin{figure}
	\center{\includegraphics[scale=0.36]{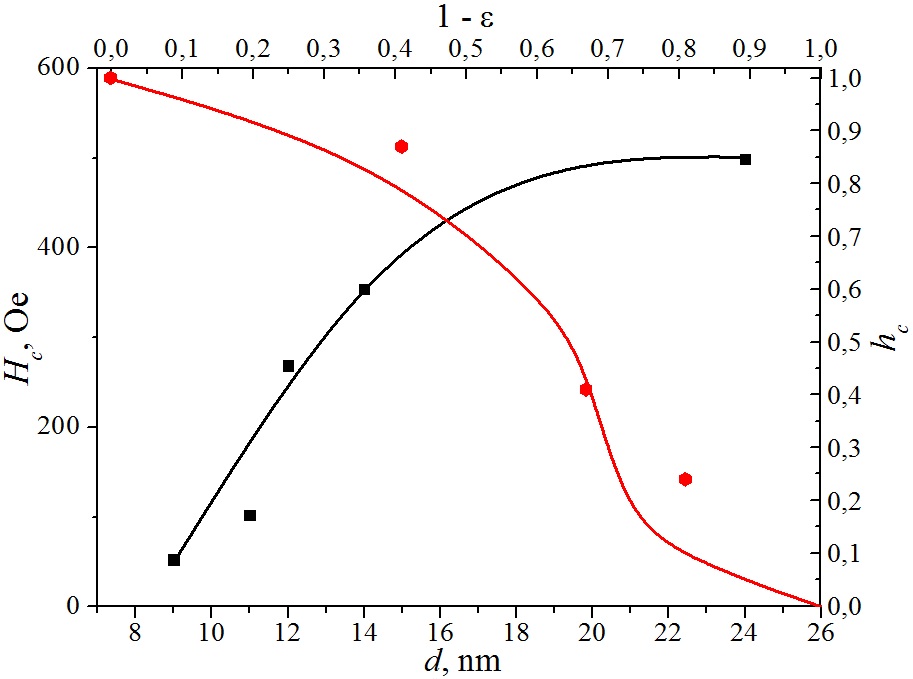}}
	\caption{Black line – dependence of the coercive field $H_c$ on the size $d$; red line – dependence of the normalized coercive field $h_c=H_c/H_{c max}$ on the relative volume of the shell $(1-\varepsilon)$ of the Fe/Fe$_3$O$_4$ nanoparticles. Black dots show the experimental results from reference\cite{Kaur}, red dots the experimental results from reference\cite{Zeng2}.}
	\label{fig:fig8}
\end{figure}

\begin{figure}
	\center{\includegraphics[scale=0.8]{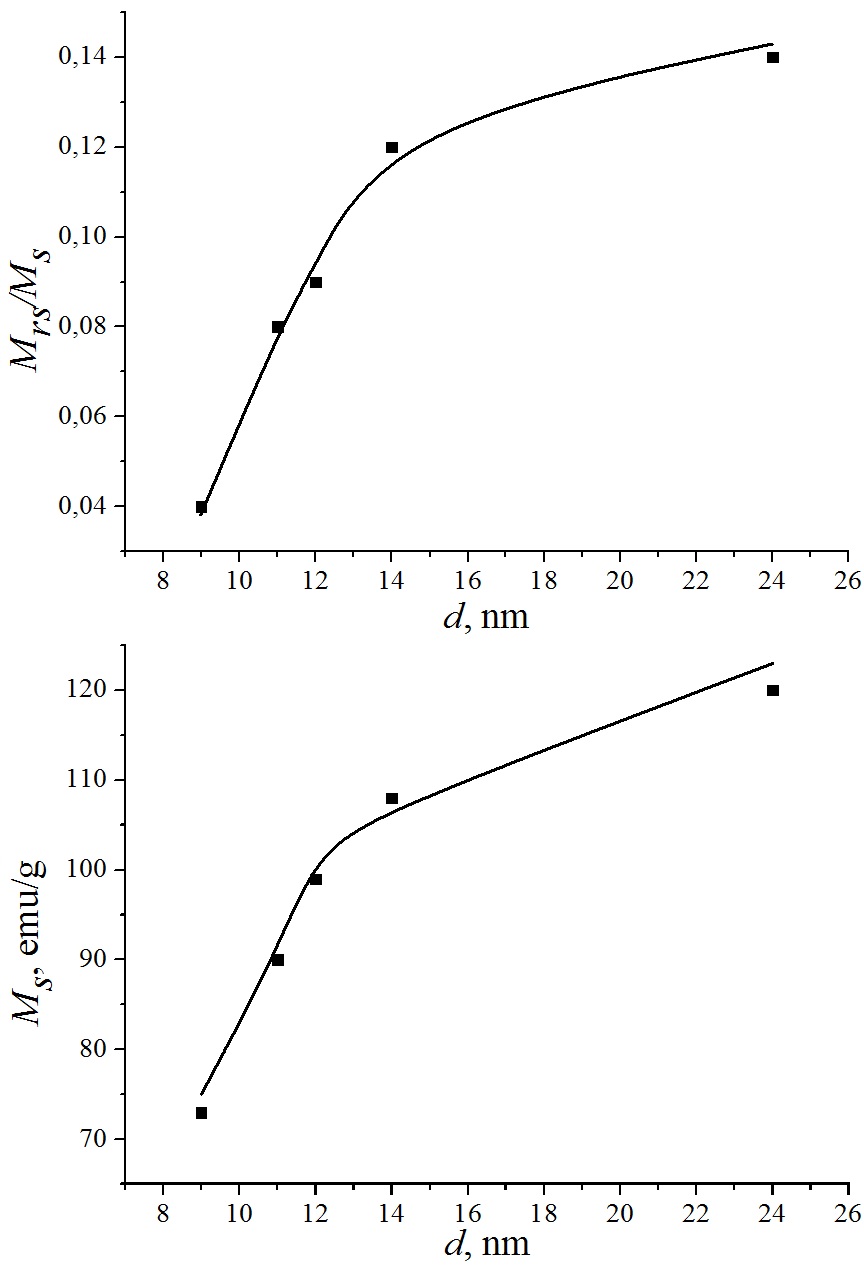}}
	\caption{a) Dependence of the $M_{rs}/M_s$ ratio on the size of the core/shell nanoparticles; b) dependence of the saturation magnetization on the size of the core/shell nanoparticles. Dots show the experimental results\cite{Kaur}, lines – theoretical data.}
	\label{fig:fig9}
\end{figure}

The results of studying the effect of their size on the hysteresis characteristics of Fe/Fe$_3$O$_4$ interacting core/shell nanoparticles are shown on the fig. \ref{fig:fig8} and \ref{fig:fig9}.

For the comparison with the results shown in the paper \cite{Kaur} we selected Fe/Fe$_3$O$_4$ nanoparticles with the values of $A_{in}$, shown in Table \ref{tabl:tabl1}.

\begin{table}
	\caption{Values of the interfacial exchange interaction constant $A_{in}$ of Fe/Fe$_3$O$_4$ nanoparticles of different sizes.}
	\label{tabl:tabl1}
	\begin{tabular} {lccccc}\hline\hline
		$ B, nm $ & 9 & 11 & 12 & 14 & 24 \\ 
		$ A_{in}, 10^{-8} erg/cm $ & -0.36 & -0.32  & -0.3 & -0.17 &-0.06 \\ 
		\hline\hline
	\end{tabular}
\end{table}

The increase of $H_c$ and $M_{rs}$ with the growing size of the nanoparticles is due to an increase of the potential barriers and a decrease of the “destructive role” of thermal fluctuations. 

For comparison with the results shown in paper \cite{Zeng2}, in fig. \ref{fig:fig8} we have shown the dependence of $h_c=H_c/H_{c max}$ on the relative volume of the shell $(1-\varepsilon)$ of $Fe/Fe_3O_4$ nanoparticles.

Figures \ref{fig:fig8} and \ref{fig:fig9} demonstrate good agreement of the theoretical data with the experiment.

\subsection{Conclusion}

In this work, the model of two-phase nanoparticles \cite{Afremov3} gets developed further: 

\begin{itemize}
	\item The previous restrictions on the anisotropy axes orientations of the core and shell and on the different types of internal interactions of each core/shell nanoparticle in the system have been removed;
	\item Interparticle interaction has been added.
\end{itemize}

A study of the spectrum of ground and metastable states, and also the size effect, the interfacial exchange interaction between core and shell, and the interparticle interaction on the hysteresis characteristics of Fe/Fe$_3$O$_4$ nanoparticles has been carried out within our model of interacting core/shell nanoparticles. In this study, it has been shown that:

\begin{itemize}
	\item In the approach of magnetic uniaxiality of the multiaxial crystal\cite{Afremov1,Afremov2}, mutual arrangements of the anisotropy axes of the core and shell have effect only on the orientation of their magnetic moments and do not change the structure of the equilibrium states;
	\item The hysteresis characteristics of a system of Fe/Fe$_3$O$_4$ nanoparticles increase with increasing nanoparticle size;
	\item When increasing the interfacial exchange interaction, the coercive field $H_c$ and the remanent saturation magnetization $M_{rs}$ change non-monotonically;
	\item Increasing the volume concentration of the core/shell nanoparticles and the intensity of their magnetostatic interaction results in a decrease of $H_c$ and $M_{rs}$.
\end{itemize}

The results obtained are in good agreement with the experimental data shown in references\cite{Kaur,Zeng2}.

\subsection{Appendix I}

\subsubsection{Magnetostatic energy of the ellipsoidal core/shell nanoparticle}

In accordance with the principle of superposition for the magnetic fields of the phases (uniformly magnetized nanoparticle and core), and according to reference\cite{Stavn}, the magnetostatic energy of the two-phase nanoparticle can be written as:

$$
\begin{matrix}
E_m=[\frac{1}{2}(1-2\varepsilon)(M_\textbf{s}^{(1)},\hat{N}^{(1)},M_s^{(1)})
+\frac{1}{2}\varepsilon
\\(M_s^{(1)},\hat{N}^{(2)},M_s^{(1)})+\frac{1}{2}\varepsilon(M_s^{(2)},\hat{N}^{(2)},M_s^{(2)})\\
+\varepsilon(M_s^{(2)},(\hat{N}^{(2)}-\hat{N}^{(1)})M_s^{(1)})]V
\end{matrix},\eqno (1\mathrm{AI})
$$

where $\varepsilon=\nu/V$ -- relative volume of the second phase, $M_s^{(1,2)}$ -- spontaneous magnetization vectors of both phases, $\hat{N}^{(1,2)}$ -- tensor of the demagnetizing coefficients of phases:

$$
\hat{\textbf{N}}^{(1)}\begin{pmatrix}
N_x^{(1)} & 0 & 0 \\
0 & N_x^{(1)} & 0 \\
0 & 0 & N_z^{(1)}
\end{pmatrix},
\hat{\textbf{N}}^{(2)}\begin{pmatrix}
N_{\tilde{x}}^{(2)} & 0 & 0 \\
0 & N_{\tilde{x}}^{(2)} & 0 \\
0 & 0 & N_{\tilde{z}}^{(2)}
\end{pmatrix},\eqno (2\mathrm{AI})
$$

Assuming that the main axis of the tensor of demagnetizing coefficients of the second phase $O\tilde{z}$ is located in the plane $ZOY$ and makes an angle $\alpha$ with the axis $OZ$ (fig.1), the tensor $\hat{N}^(2)$ can be written as:

$$
\hat{\textbf{N}}^{(2)}=
\begin{pmatrix}
N_{\tilde{x}}^{(2)} & 0 & 0 \\
0 & \frac{N_{\tilde{x}}^{(2)}+N_{\tilde{z}}^{(2)}+k_n^{(2)}\cos2\alpha}{2} & \frac{k_N^{(2)}}{2}\sin2\alpha \\
0 &  \frac{k_N^{(2)}}{2}\sin2\alpha & \frac{N_{\tilde{x}}^{(2)}+N_{\tilde{z}}^{(2)}-k_n^{(2)}\cos2\alpha}{2}
\end{pmatrix},\eqno (3\mathrm{AI})
$$

where $k_N^{(2)}=N_{\tilde{x}}^{(2)}-N_{\tilde{z}}^{(2)}$.

Defining $M_s^{(1,2)}$ by the projections $M_s^{(1,2)} = \{0, M_s^{(1,2)}\sin\vartheta^{(1,2)}, M_s^{(1,2)}\cos\vartheta^{(1,2)}\}$, we obtain the following equation for the magnetostatic energy:

$$
\begin{matrix}
E_m =\{-\frac{(M_s^{(1)})^2}{4}[((1-2\varepsilon)k_N^{(1)}+\\\varepsilon k_N^{(2)}\cos 2\alpha)\cos 2\vartheta^{(1)}
-\varepsilon k_N^{(2)}\sin 2\alpha\sin 2\vartheta^{(1)}]-\\\frac{(M_s^{(2)})^2}{4}\varepsilon[k_N^{(2)}\cos 2\alpha\cos 2\vartheta^{(2)}
-k_N^{(2)}\sin 2\alpha\sin 2\vartheta^{(2)}]+\\\frac{1}{3}\varepsilon M_s^{(1)}M_s^{(2)}[(k_N^{(2)}-k_N^{(1)})(\sin\vartheta^{(1)}\sin\vartheta^{(2)}
-\\2\cos\vartheta^{(1)}\cos\vartheta^{(2)})+\frac{3}{2}k_N^{(2)}\sin 2\alpha\sin (\vartheta^{(1)}+\vartheta^{(2)})]\}V,
\end{matrix}\eqno (4\mathrm{AI})
$$

Here $k_N^{(1)}=N_x^{(1)}-N_z^{(1)}$. If the long axes of the particle and core coincide, then

$$
\begin{matrix}
E_m =[-\frac{(M_s^{(1)})^2}{4}((1-2\varepsilon)k_N^{(1)}+\varepsilon k_N^{(2)})\cos 2\vartheta^{(1)}\\
-\frac{(M_s^{(2)})^2}{4}\varepsilon k_N^{(2)}\cos 2\vartheta^{(2)}+\frac{1}{3}\varepsilon M_s^{(1)}M_s^{(2)}(k_N^{(2)}-k_N^{(1)})\\
(\sin\vartheta^{(1)}\sin\vartheta^{(2)}-2\cos\vartheta^{(1)}\cos\vartheta^{(2)})]V
\end{matrix},\eqno (5\mathrm{AI})
$$

The resulting equation coincides with the equation for the magnetostatic energy, which has been obtained within the model of “plane parallel phases”\cite{Belokon,Otero}, if $
N_{11}=(1-2\varepsilon)k_N^{(1)}+\varepsilon k_N^{(2)}, N_{22}=\varepsilon k_N^{(2)},N_{12}=-N_{21}/2=\varepsilon(k_N^{(2)}-k_N^{(1)})/3.$

\subsection{Appendix II}

When the long axes coincide with each other ($\alpha=0$), studying the energy (\ref{eq:eq5}) for an extreme point allows us to calculate the spectrum of the potential barriers $E_{ik}$, separating the equilibrium states of the two-phase nanoparticle, situated in the external field $H$ parallel to the axis $Oz$:

$$
E_{12}=E_{34}=\frac{\varepsilon^{2}(H_c^{(1\to 2)}+H)^{2}}{2\mathcal{K}^{(2)}}V,\eqno (1\mathrm{AII})
$$

$$
E_{21}=E_{43}=\frac{\varepsilon^{2}(H_c^{(2\to 1)}-H)^{2}}{2\mathcal{K}^{(2)}}V,\eqno (2\mathrm{AII})
$$

$$
E_{13}=\frac{(H_c^{(1\to 3)}+H)^2M_s^{(\uparrow\uparrow)}}{2H_c^{(1\to 3)}}V,\eqno (3\mathrm{AII})
$$

$$
E_{31}=\frac{(H_c^{(3\to 1)}-H)^2M_s^{(\uparrow\uparrow)}}{2H_c^{(3\to 1)}}V,\eqno (4\mathrm{AII})
$$

$$
E_{14}=\frac{(1-\varepsilon)^2(H_c^{(1\to 4)}+H)^2}{2\mathcal{K}^{(1)}}V,\eqno (5\mathrm{AII})
$$

$$
E_{41}=\frac{(1-\varepsilon)^2(H_c^{(4\to 1)}-H)^2}{2\mathcal{K}^{(1)}}V,\eqno (6\mathrm{AII})
$$

$$
E_{23}=\frac{(1-\varepsilon)^2(H_c^{(2\to 3)}+H)^2}{2\mathcal{K}^{(1)}}V,\eqno (7\mathrm{AII})
$$

$$
E_{32}=\frac{(1-\varepsilon)^2(H_c^{(3\to 2)}-H)^2}{2\mathcal{K}^{(1)}}V,\eqno (8\mathrm{AII})
$$

$$
E_{24}=\frac{(H_c^{(2\to 4)}+H)^2M_s^{(\uparrow\downarrow)}}{2H_c^{(2\to 4)}}V,\eqno (9\mathrm{AII})
$$

$$
E_{42}=\frac{(H_c^{(4\to 2)}-H)^2M_s^{(\uparrow\downarrow)}}{2H_c^{(4\to 2)}}V,\eqno (10\mathrm{AII})
$$

Here $M_s^{(\uparrow\uparrow)}=\left( (1-\varepsilon)M_s^{(1)}+\varepsilon M_s^{(2)}\right)$, $M_s^{(\uparrow\downarrow)}=|(1-\varepsilon)M_s^{(1)}-\varepsilon M_s^{(2)}|$, $H_c^{(i\rightarrow k)}$ -- critical field of the transition from $i$-th to $k$-th state\cite{Afremov2} are defined by geometric and magnetic characteristics of the core/shell nanoparticle:

$$
H_c^{(1\to 2)}=H_c^{(3\to 4)}=\frac{|\mathcal{K}^{(2)}|M_s^{(2)}-\mathcal{U}_{2}M_s^{(1)}}{\varepsilon},\eqno (11\mathrm{AII})
$$

$$
\begin{matrix}
H_c^{(1\to 3)}=H_c^{(3\to 1)}=\frac{1}{M_s^{(\uparrow\uparrow)}}|\mathcal{K}^{(1)}|(M_s^{(1)})^2+\\|\mathcal{K}^{(2)}|(M_s^{(2)})^2+2(k_N^{(2)}-k_N^{(1)})M_s^{(1)}M_s^{(2)}
\end{matrix},\eqno (12\mathrm{AII})
$$

$$
H_c^{(1\to 4)}=H_c^{(3\to 2)}=\frac{|\mathcal{K}^{(1)}|M_s^{(1)}-\mathcal{U}_2M_s^{(2)}}{1-\varepsilon},\eqno (13\mathrm{AII})
$$

$$
H_c^{(2\to 1)}=H_c^{(4\to 3)}=\frac{|\mathcal{K}^{(2)}|M_s^{(2)}+\mathcal{U}_2M_s^{(1)}}{\varepsilon},\eqno (14\mathrm{AII})
$$

$$
\begin{matrix}
H_c^{(2\to 4)}=H_c^{(4\to 2)}=\frac{1}{|M_s^{(\downarrow\uparrow)}|}|\mathcal{K}^{(1)}|(M_s^{(1)})^2+\\|\mathcal{K}^{(2)}|(M_s^{(1)})^2-2(k_N^{(2)}-k_N^{(1)})M_s^{(1)}M_s^{(2)}
\end{matrix},\eqno (15\mathrm{AII})
$$

$$
H_c^{(4\to 1)}=H_c^{(2\to 3)}=\frac{|\mathcal{K}^{(1)}|M_s^{(1)}+\mathcal{U}_2M_s^{(2)}}{1-\varepsilon},\eqno (16\mathrm{AII})
$$

\end{document}